# Cylindrical Ising Nanowire in an Oscillating Magnetic Field and Dynamic Compensation Temperature

**Ersin Kantar and Mehmet Ertaş[1]**

*Department of Physics, Erciyes University, 38039 Kayseri, Turkey*

**Abstract** The magnetic properties of a nonequilibrium spin-1/2 cylindrical Ising nanowire system with core/shell in an oscillating magnetic field are studied by using a mean-field approach based on the Glauber-type stochastic dynamics (DMFT). We employ the Glauber-type stochastic dynamics to construct set of the coupled mean-field dynamic equations. First, we study the temperature dependence of the dynamic order parameters to characterize the nature of the phase transitions and to obtain the dynamic phase transition points. Then, we investigate the temperature dependence of the total magnetization to find the dynamic compensation points as well as to determine the type of behavior. The phase diagrams in which contain the paramagnetic, ferromagnetic, antiferromagnetic, nonmagnetic, surface fundamental phases and tree mixed phases as well as reentrant behavior are presented in the reduced magnetic field amplitude and reduced temperature plane. According to values of Hamiltonian parameters, the compensation temperatures, or the N-, Q-, P-, R-, S-type behaviors in the Néel classification nomenclature exist in the system.

## 1. Introduction

In recent four years, the phenomenon of magnetic nanostructures with a fascinating variety of morphologies (nanoparticle, nanotube and nanowire) has been one of the intensively studied subjects in statistical mechanics and condensed matter physics, because of their potential device applications in technologically important materials (see [1-6] and references therein). These systems can find application in fields such as ensure super high data storage densities, sensors, biomedicine and catalysis, among others [7].

On the other hand, many researchers have used the spin-1/2 Ising system to study equilibrium properties of magnetic nanostructured materials (see [8-14] and references therein). An early attempt to examine nanoparticles was done by Kaneyoshi [8]. In this study, the phase diagrams of a ferroelectric nanoparticle described with the transverse Ising model were investigated by using the mean-field theory (MFT) and the effective-field theory (EFT). MFT and EFT were used to study the magnetizations [9] and phase diagrams [10] of a transverse Ising nanowire and found that the equilibrium behavior of the system is strongly affected by the surface situations. Total susceptibility, susceptibility, compensation temperature and phase diagrams of a cylindrical spin-1/2 Ising nanotube (or nanowire) were examined by utilizing the EFT in detail [11]. The cylindrical nanowire system with a diluted surface described by the transverse spin-1/2 Ising model was investigated by using the EFT [12]. Akıncı [13] examined the effects of the randomly distributed magnetic field on the phase diagrams of a spin-1/2 Ising nanowire with the EFT. Zaim et al. [14] studied the hysteresis behaviors of the nanotube in which consisting of a ferroelectric core of spin-1/2 surrounded. Furthermore, although a great amount the spin-1/2 Ising systems have used to investigate the equilibrium properties of magnetic nanostructured materials, there have been only a few works that the spin-1/2 Ising systems used to investigate dynamic magnetic properties of nanostructured materials [15-20]. In series of these works, effective-field theory based on the Glauber-type stochastic (DEFT) was used as method.

[1] Corresponding author. E-mail: mehmetertas@erciyes.edu.tr (Mehmet Ertaş)

We also mention that dynamic compensation temperatures (DCTs) and the dynamic phase transition (DPT) temperatures have been attracted much attention (see [21-28] and references therein) in recent years. The existence of compensation temperatures is of great technological importance, since at this point only a small driving field is required to change the sign of the total magnetization. This property is very useful in thermomagnetic recording, electronic, and computer technologies. On the other hand, experimental realizations for the DPT have been discussed in ultrathin ferromagnetic films, superconductors, ferroicsystems, ultrathin polyethylene naphthalate nanocomposites, Al–Ni systems, etc. (see [29–32] and references therein).

As far as we know, the dynamic compensation behaviors of the nano systems (nanoparticle, nanotube and nanowire) have been studied only two works [16, 19]. In two works, the compensation types of the system were found by using DEFT. The dynamic phase diagrams of the system were not investigated. Moreover, the dynamic behaviors of the nano system have not been examined by using a mean-field approach based on the Glauber-type stochastic dynamics (DMFT). Therefore, in present paper, we used to DMFT for study the magnetic properties of a nonequilibrium spin-1/2 cylindrical Ising nanowire system with core/shell in an oscillating magnetic field. The aim of the present paper is three-fold: (i) to obtain the DCTs temperatures and DPT of the Ising nanowire system. (ii) To investigate the type of the compensation behavior of the system. (iii) Finally, to present the dynamic phase diagrams of the system in the plane of temperature versus magnetic field amplitude.

The outline of the rest of the paper is follows. In Sec. 2, the model is presented briefly and the derivation of the mean-field dynamic equations of motion is given by using the Glauber-type stochastic dynamics in the presence of an oscillating magnetic field. In Sec. 3, the numeric results and discussions are presented. Finally, we give the summary and conclusion in the last section.

## 2. Model and Formulations

The considered model is a spin-1/2 Ising nanowire on a cylindrical lattice under the oscillating magnetic field. The schematic representation of a cylindrical Ising nanowire is depicted in Fig. 1, in which the wire consists of the surface shell and the core. Each site on the figure is occupied by a spin-1/2 Ising particle and each spin is connected to the two nearest-neighbor spins on the above and below sections along the cylinder. The Hamiltonian of the system is given by

$$\begin{aligned}\mathcal{H} = & -J_C\left(\sum_{ii'}\sigma_i\sigma_{i'}+\sum_{ij}\sigma_i S_j+\sum_{jj'}S_j S_{j'}\right)-J_1\left(\sum_{jk}S_j\alpha_k+\sum_{jl}S_j\lambda_l\right)\\ & -J_S\left(\sum_{kk'}\alpha_k\alpha_{k'}+\sum_{kl}\alpha_k\lambda_l+\sum_{ll'}\lambda_l\lambda_{l'}\right)-H\left(\sum_i\sigma_i+\sum_j S_j+\sum_k\alpha_k+\sum_l\lambda_l\right)\end{aligned} \quad (1)$$

where the $J_S$ and $J_C$ are the exchange interaction parameters between two nearest-neighbor magnetic particles at the surface shell and core, respectively, and $J_1$ is the interaction parameters between two nearest-neighbor magnetic particles at the surface shell and the core shell. The surface exchange and interfacial coupling interactions are often defined as $J_S = J_C(1+\Delta_S)$ and $r = J_1/J_C$ in the nanosystems [9, 10, 33, 34], respectively. H is the oscillating magnetic field: $H(t) = H_0 \cos(wt)$, with $H_0$ and $w = 2\pi\nu$ being the amplitude and the angular frequency of the oscillating field, respectively. The system is in contact with an isothermal heat bath at an absolute temperature $T_A$.

Now, we apply the Glauber-type stochastic dynamics to obtain the set of the mean-field dynamic equations. Since the derivation of the mean-field dynamic equations was described in detail for spin-1/2 system [35] and different spin systems [21-24], in here, we shall only give a brief summary. If the $S, \alpha$ and $\lambda$ spins momentarily fixed, the master equation for σ- spins can be written as

$$\frac{d}{dt}P(\sigma_1,\sigma_2,...,\sigma_N;t) = -\sum_i W_i(-\sigma_i)P(\sigma_1,\sigma_2,...,\sigma_i,...\sigma_N;t) + \sum_i W_i(\sigma_i)P(\sigma_1,\sigma_2,...,-\sigma_i,...\sigma_N;t), \tag{2}$$

where $W_i(\sigma_i)$ is the probability per unit time that *i*th σ spin changes from $\sigma_i$ to $-\sigma_i$. Since the system is in contact with a heat bath at absolute temperature $T_A$, each spin σ can flip with the probability per unit time by the Boltzmann factor;

$$W_i(\sigma_i) = \frac{1}{\tau}\frac{\exp(-\beta\,\Delta E(\sigma_i))}{\sum_{\sigma_i}\exp(-\beta\,\Delta E(\sigma_i))}, \tag{3}$$

where $\beta = 1/k_B T_A$, $k_B$ is the Boltzmann constant, the sum ranges the two possible values ±1/2 for $\sigma_i$ and

$$\Delta E(\sigma_i) = 2\sigma_i (H + z_{\sigma\sigma} J_C \sum_{i'} \sigma_{i'} + z_{\sigma S} J_C \sum_j S_j), \tag{4}$$

gives the change in the energy of the system when the $\sigma_i$-spin changes. The probabilities satisfy the detailed balance condition. Using Eqs. (2), (3), (4) with the mean-field approach, we obtain the mean-field dynamic equation for the σ-spins as

$$\Omega \frac{d}{d\xi} m_{C1} = -m_{C1} + \frac{1}{2}\tanh\left[\beta\left(z_{\sigma\sigma} J_C m_{C1} + z_{\sigma S} J_C m_{C2} + H_0 \cos(\xi)\right)\right], \tag{5}$$

where $m_{C1} = \langle\sigma\rangle$, $m_{C2} = \langle S\rangle$, $\xi = wt$ and $\Omega = \tau w = w/f$, w is the frequency of the oscillating magnetic field and $f$ represents the frequency of spin flipping. Moreover, $z_{\sigma\sigma}$ and $z_{\sigma S}$ corresponds to the number of nearest-neighbor pairs of spins σ−σ and σ−S, respectively, in which $z_{\sigma\sigma} = 2$ and $z_{\sigma S} = 6$.

As similar to σ-spins, we obtain the mean-field dynamical equations for the $S, \alpha$ and $\lambda$ -spins by using the similar calculations. The mean-field dynamic equations for $S, \alpha$ and $\lambda$ -spins are obtained as

$$\Omega \frac{d}{d\xi} m_{C2} = -m_{C2} + \frac{1}{2} \tanh\left[\beta\left(z_{S\sigma} J_C m_{C1} + z_{SS} J_C m_{C2} + z_{S\alpha} J_1 m_{S1} + z_{s\lambda} J_1 m_{S2} + H_0 \cos\left(\xi\right)\right)\right], \quad (6)$$

$$\Omega \frac{d}{d\xi} m_{S1} = -m_{S1} + \frac{1}{2} \tanh\left[\beta\left(z_{\alpha\alpha} J_S m_{S1} + z_{\alpha\lambda} J_S m_{S2} + z_{\alpha S} J_1 m_{C2} + H_0 \cos\left(\xi\right)\right)\right], \quad (7)$$

$$\Omega \frac{d}{d\xi} m_{S2} = -m_{S2} + \frac{1}{2} \tanh\left[\beta\left(z_{\lambda\lambda} J_S m_{S2} + z_{\lambda\alpha} J_S m_{S1} + z_{\lambda S} J_1 m_{C2} + H_0 \cos\left(\xi\right)\right)\right], \quad (8)$$

where $m_{S1} = \langle\alpha\rangle$, $m_{S2} = \langle\lambda\rangle$, $z_{S\sigma} = z_{S\alpha} = z_{\alpha S} = 1$, $z_{SS} = 4$, $z_{s\lambda} = z_{\alpha\alpha} = z_{\alpha\lambda} = z_{\lambda\lambda} = z_{\lambda\alpha} = z_{\lambda S} = 2$. Hence, a set of mean-field dynamical equations of the system are obtained. We fixed $J_C = 1$ that the core shell interaction is ferromagnetic and $\Omega = 2\pi$.

We should also mention that the dynamic compensation temperature, which dynamic total magnetization ($M_t$) vanishes at the compensation temperature $T_{Comp}$. The compensation point can then be determined by looking for the crossing point between the absolute values of the surface and the core magnetizations. Therefore, at the compensation point, we must have

$$\left|M_{Surface}\left(T_{Comp}\right)\right| = \left|M_{Core}\left(T_{Comp}\right)\right|, \quad (9)$$

and

$$\operatorname{sgn}\left[M_{Surface}\left(T_{Comp}\right)\right] = -\operatorname{sgn}\left[M_{Core}\left(T_{Comp}\right)\right]. \quad (10)$$

We also require that $T_{Comp} < T_C$, where $T_C$ is the critical point temperature. In the next section we will give the numerical results of these dynamic equations.

**3. Numerical Results and Discussions**

In this section, we solved first the Eqs. (5)-(8) to find the phases in the system. These equations were solved by using the numerical method of the Adams-Moulton predictor corrector method and we found that a paramagnetic (P), ferromagnetic (F), antiferromagnetic (AF), nonmagnetic (NM), surface phase (SF) fundamental phases and four mixed phases, namely the F + P in which F, P phases coexist, the F + AF in which F, AF phases coexist, the AF + P in which AF, P phases coexist, and the NM+P in which NM, P phases coexist, exist in the system. Since we gave the solution of these kinds of dynamic equations in Ref. 22-24, 27 in detail, we will not discuss the solutions and present any figures here.

Then, we investigate the behavior of the dynamic core and shell magnetizations ($M_{C1}$, $M_{C2}$, $M_{S1}$, $M_{S2}$) as a function of the temperature. This investigation leads us to characterize the nature (continuous or discontinuous) of phase transitions and find the DPT points. Furthermore, we study the dynamic total magnetization as a function of the temperature to find the dynamic compensation temperatures and to determine the type of behavior. The dynamic shell and core magnetizations and the dynamic total magnetizations $M_t = \left(m_{C1} + 6m_{C2} + 6\left(m_{S1} + m_{S2}\right)\right)/19$ as the time-averaged magnetization over a period of the oscillating magnetic field are given as

$$M_{C1,C2,S1,S2} = \frac{1}{2\pi}\int_0^{2\pi} m_{C1,C2,S1,S2}(\xi)\, d\xi \,, \tag{11a}$$

and

$$M_t = \frac{1}{2\pi}\int_0^{2\pi} \left(\frac{m_{C1}(\xi) + 6m_{C2}(\xi) + 6\left(m_{S1}(\xi) + m_{S2}(\xi)\right)}{19}\right) d\xi \,. \tag{11b}$$

The behaviors of $M_{C1,}$ $M_{C2}$, $M_{S1,}$ $M_{S2}$ and $M_t$ as functions of the temperature for several values of interaction parameters are obtained by solving Eqs. 11(a) and (b). In order to show the calculation of the DPT points and the compensation temperatures, the three explanatory and interesting examples are plotted in Figs. 2(a)-(c). In these figures, $T_C$ and $T_t$ are the second-order and first-order phase transition temperatures, respectively. $T_{Comp}$ is the compensation temperature. Fig. 2(a) shows the behavior of $M_{C1,}$ $M_{C2}$, $M_{S1,}$ $M_{S2}$ and $M_t$ as functions of the temperature for $r = -0.5$, $\Delta_S = -0.6$ and $H_0$ = 0.1. $M_{C1,\ C2}$ = 0.5 and $M_{S1,\ S2}$ = -0.5 are at the zero temperature. The dynamic magnetizations $M_{C1}$, $M_{C2}$ decrease and $M_{S1,}$ $M_{S2}$ increase continuously with the increasing of the values of temperature below the critical temperature and they become zero at $T_C$ = 2.95; hence, a second-order phase transition occurs. The transition is from the AF phase to the P phase. Moreover, one compensation temperature or N-type behavior occurs in the system that exhibits the same behavior classified after the Néel theory [36] as the N-type behavior [37]. In Fig. 2(b), $M_{C1,\ C2,\ S1,\ S2}$ = 0.5 at the zero temperature; thus we own the F phase; as the temperature increases, all of them decrease to zero continuously and the system undergoes a second-order phase transition from the F phase to the P phase at $T_C$ = 3.44. In Fig. 2(c), at zero value of temperature, $M_{C1,\ C2}$ = 0.5 and $M_{S1,\ S2}$ = 0.0; thus we have the NM phase; as the temperature increases, $M_{C1,\ C2}$ of them decrease to zero discontinuously and the system undergoes a first-order phase transition from the NM phase to the P phase at $T_t$ = 0.31.

On the other hand, one of the typical dynamic behaviors in such a system is to show a compensation point below its transition temperature. We studied dynamic compensation behaviors of the cylindrical Ising nanowire system, which we know that a compensation temperature can be found in the nanostructure systems [16, 19]. Fig. 3 shows the temperature dependencies of the total magnetization $M_t$ for H=0.1 and several values of r, $\Delta_S$. As seen from Fig. 3 (a), the curve labeled $r = -1.0$ and $\Delta_S = -0.5$ may show the N-type behavior. Moreover, the Q-type behavior is obtained in Fig. 3 (b) for $r = -0.5$ and $\Delta_S = 0.5$. Fig. 3(c) is calculated for $r = -0.1$ and $\Delta_S = 1.0$. As we can see this curve illustrates the P-type behavior. Fig. 3(d) is calculated for $r = 1.0$ and $\Delta_S = -0.75$ and illustrates the R-type behavior. For $r = 1.0$ and $\Delta_S = -0.99$, Fig. 3(e) is obtained and exhibit the S-type behavior. These obtained results are consistent with same behavior as that classified in the Neél theory [36].

Since we have obtained DPT points and compensation temperatures, we can now present the dynamic phase diagrams of the system. The calculated phase diagrams for both antiferromagnetic case and ferromagnetic case in the (T, h) plane is presented in Fig. 4 and in Fig. 5, respectively. In these dynamic phase diagrams, the solid, dashed and dash-dot-dot lines represent the second-order, first-order phase transitions temperatures and the compensation temperatures, respectively. The dynamic tricritical point is denoted by a filled circle. Moreover, z, tp, and qp are the dynamic zero temperature, triple and quadruple points, respectively, that strongly depend on the values of interaction parameters.

Fig. 4 shows the dynamic phase diagrams including the compensation behaviors for antiferromagnetic case in (T, h) plane and four main topological different types of phase diagrams are seen. From these phase diagrams, the following phenomena have been observed. (i) Figs. 4(a)-(c) include the compensation temperatures, but Fig. 4(g) does not. (ii) The dynamic phase diagrams contain the F, AF, NM, SF fundamental phases as well as the F + AF, AF + P, NM + P mixed phases. (iii) The system exhibits the reentrant behavior, i.e., with the temperature increase, the system passes from the P phase to the AF phase, and then back to the P phase again in seen Fig. 4(c). It should also mentiononig that several weakly frustrated ferromagnets, such as in manganite $LaSr_2Mn_2O_7$ by electron and x-ray diffraction, in the bulk bicrystals of the oxide superconductor $BaPb_{1-x}Bi_xO_3$ and $Eu_xSr_{1-x}S$ and amorphous-$Fe_{1-x}Mn_x$, demonstrate the reentrant phenomena [37-40]. (iv) The system shows qp, where the four first-order transition lines meet, is shown in Fig. 4(c). tp, at which three first-order transition lines meet in seen Figs. 4(c) and 4(d). z, in which is a critical point characterized by fluctuations at zero-temperature, is shown Fig. 4(d).

Fig. 5 exhibits the dynamic phase diagrams for ferromagnetic case in (T, h) plane and three main topological different types of phase diagrams are seen. From these phase diagrams, the following phenomena have been observed. (i) These dynamic phase diagrams do not contain the compensation temperature. (ii) The phase diagrams show one or two dynamic tricritical points. (iii) The system contains the P, F, NM, SF fundamental phases as well as the F + P, NM + P mixed phases. (iv) The system shows re-entrant behavior, seen in Fig. 5(b). (v) The dynamic phase boundaries among the fundamental phases are mostly second-order phase transition lines but between the fundamental and mixed phases are first-order phase transition lines.

## 4. Summary and Conclusion

In this work, we have studied the magnetic properties of a nonequilibrium spin-1/2 cylindrical Ising nanowire system with core/shell in an oscillating magnetic field by using a mean-field approach based on the Glauber-type stochastic dynamics (DMFT). We employ the Glauber-type stochastic dynamics to construct set of the coupled mean-field dynamic equations. First, we study the temperature dependence of the dynamic order parameters to characterize the nature of the phase transitions and to obtain the dynamic phase transition points. Then, we investigate the temperature dependence of the total magnetization to find the dynamic compensation points as well as to determine the type of behavior. The phase diagrams in which contain the paramagnetic, ferromagnetic, antiferromagnetic, nonmagnetic, surface fundamental phases and tree mixed phases as well as reentrant behavior are presented in the reduced magnetic field amplitude and reduced temperature plane. The phase diagrams also include one or two dynamic tricritical points as well as the tp, qp, z special points. According to values of Hamiltonian parameters, the system the compensation temperatures, or the N-, P-, Q-, S-, R-type behaviors in the Néel classification nomenclature exist in the system.

We should also mention that the findings of this study have important for researchers in statistical mechanics and condensed matter physics. Because, up to this time, the dynamic behaviors of nanostructure (nanoparticle, nanowire, nanotube etc.) were only studied for spin-1/2

Ising system by using an effective-field theory based on the Glauber-type stochastic dynamics (DEFT). Unfortunately, the dynamic behavior of nanostructure with high spin systems by using the DEFT due to difficulties in implementation to high spin systems of the method were not investigated. Therefore, we hope that the presented paper may be pioneer for researchers who want to examine dynamic behavior of high spin nanostructure. Lastly, we also hope this study will contribute to the theoretical and experimental research on the dynamic magnetic properties of nanostructure Ising systems as well as to research on magnetism.

**List of the figure captions**

**Fig. 1. (color online)** Schematic representations of a cylindrical nanowire: (a) cross-section and (b) three-dimensional. The gray and blue circles indicate spin-1/2 Ising particles at the surface shell and core, respectively. (For interpretation of the references to color in this figure legend, the reader is referred to the web version of this article).

**Fig. 2. (color online)** The temperature dependence of the dynamic core and shell magnetizations and total magnetizations. $T_C$ and $T_t$ are the second-order and first-order phase transition temperatures, respectively. $T_{comp}$ is the compensation temperature. Dash-dot-dot lines represent the compensation temperatures,

- **(a)** Exhibiting a second-order phase transition from the AF phase to P phase at $T_C$ = 2.95 for $r=-0.5, \Delta_S=-0.6$ and $H_0$ = 0.1. The system shows the N-type compensation behavior.
- **(b)** Exhibiting a second-order phase transition from the F phase to P phase at $T_C$ = 3.44 for $r=1.0, \Delta_S=0.0$ and $H_0$ = 0.1.
- **(c)** Exhibiting a first-order phase transition from the NM phase to P phase at $T_t$ = 0.31 for $r=-0.5, \Delta_S=-0.6$ and $H_0$ = 2.88.

**Fig. 3.** The dynamic total magnetization as a function of the temperature for different values of interaction parameters. The system exhibits the N-, Q-, P-, R-, S- type behaviors of compensation behaviors. **(a)** $r= -1.0$ and $\Delta_S=-0.5$; **(b)** $r= -0.5$ and $\Delta_S=0.5$; **(c)** $r= -0.1$ and $\Delta_S=1.0$; **(d)** $r= 1.0$ and $\Delta_S=-0.75$; **(e)** $r= 1.0$ and $\Delta_S=-0.99$.

**Fig. 4.** The dynamic phase diagrams for antiferroagnetic case. Dashed and solid lines are the dynamic first- and second-order phase boundaries, respectively. The dash-dot-dot line illustrates the compensation temperatures. The dynamic tricritical points are indicated with filled circles. **(a)** $r=-0.5$ and $\Delta_S=-0.6$; **(b)** $r=-1.0$ and $\Delta_S=0.0$; **(c)** $r=-0.05$ and $\Delta_S=0.0$; **(d)** $r=-0.2$ and $\Delta_S=-5.0$.

**Fig. 5.** Same as Fig 4, but **(a)** $r=0.2$ and $\Delta_S=-5.0$; **(b)** $r=1.0$ and $\Delta_S=3.0$; **(c)** $r=1.0$ and $\Delta_S=0.0$.

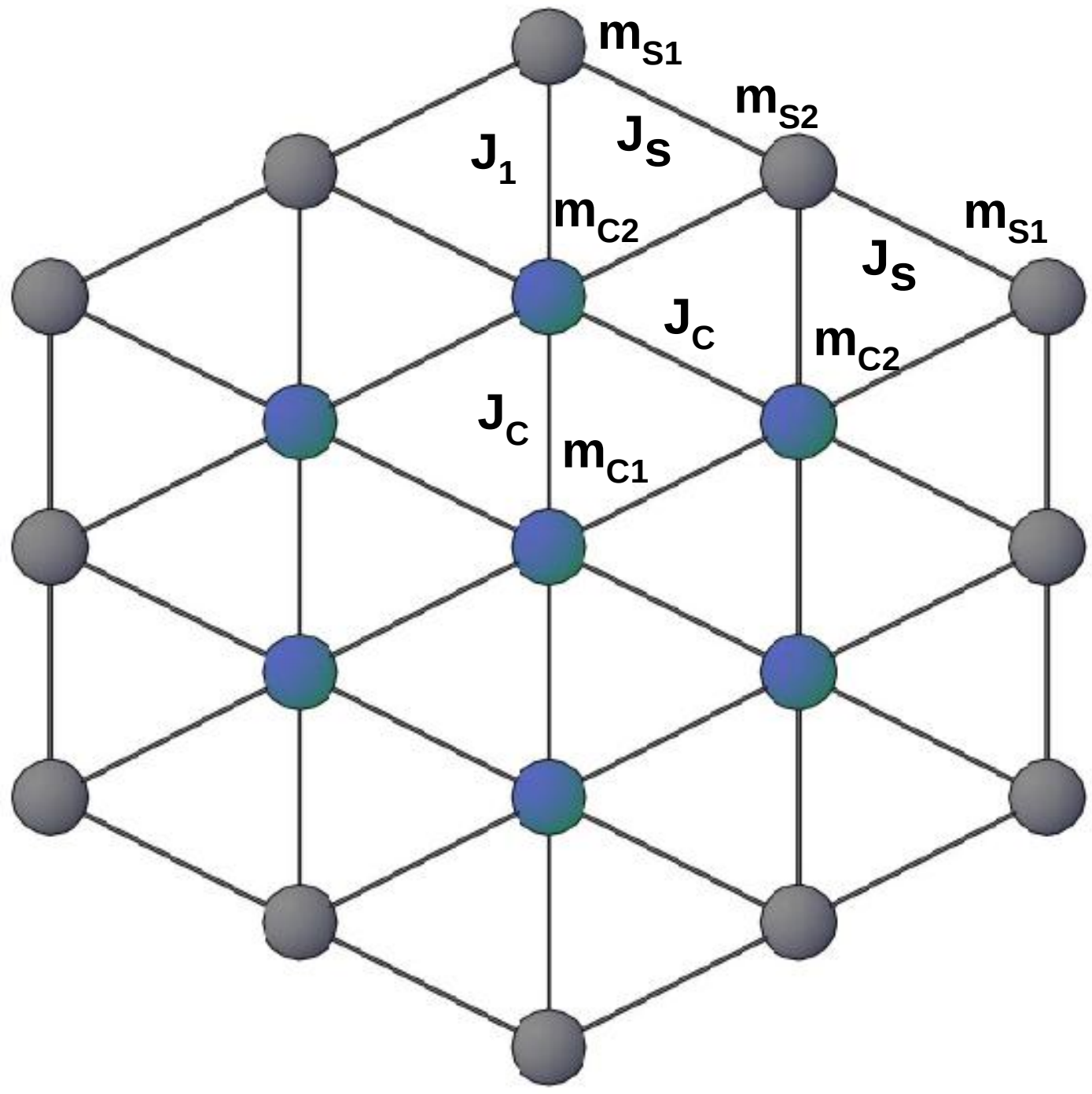

(a)

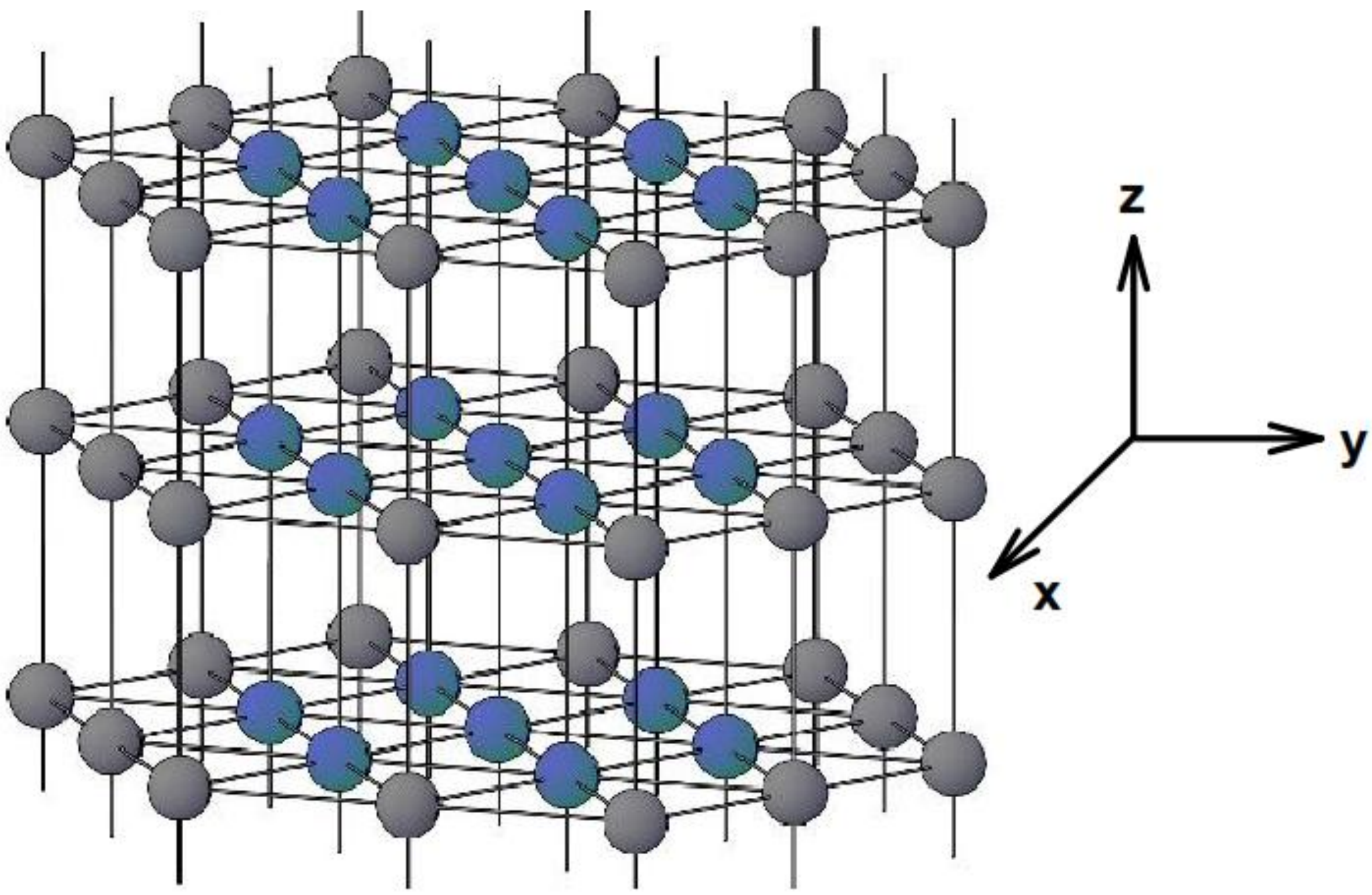

(b)

Fig. 1

Fig. 2

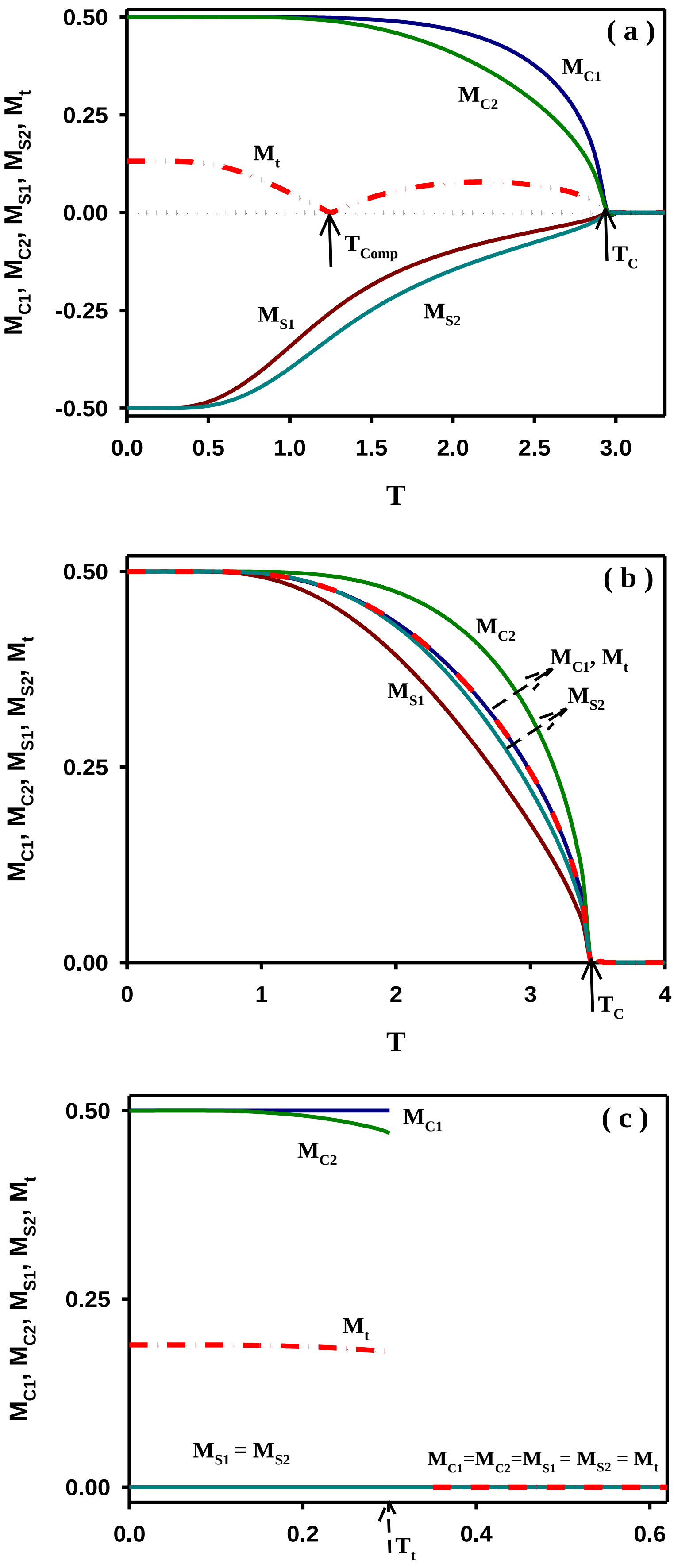

( a )
0.50
0.25
0.00
-0.25
-0.50
M_C1, M_C2, M_S1, M_S2, M_t
M_C1
M_C2
M_t
T_Comp
T_C
M_S1
M_S2
0.0
0.5
1.0
1.5
2.0
2.5
3.0
T
( b )
M_C2
M_C1, M_t
M_S1
M_S2
T_C
0
1
2
3
4
T
( c )
M_C1
M_C2
M_t
M_S1 = M_S2
M_C1=M_C2=M_S1 = M_S2 = M_t
T_t
0.0
0.2
0.4
0.6
T

Fig. 2

Fig. 3

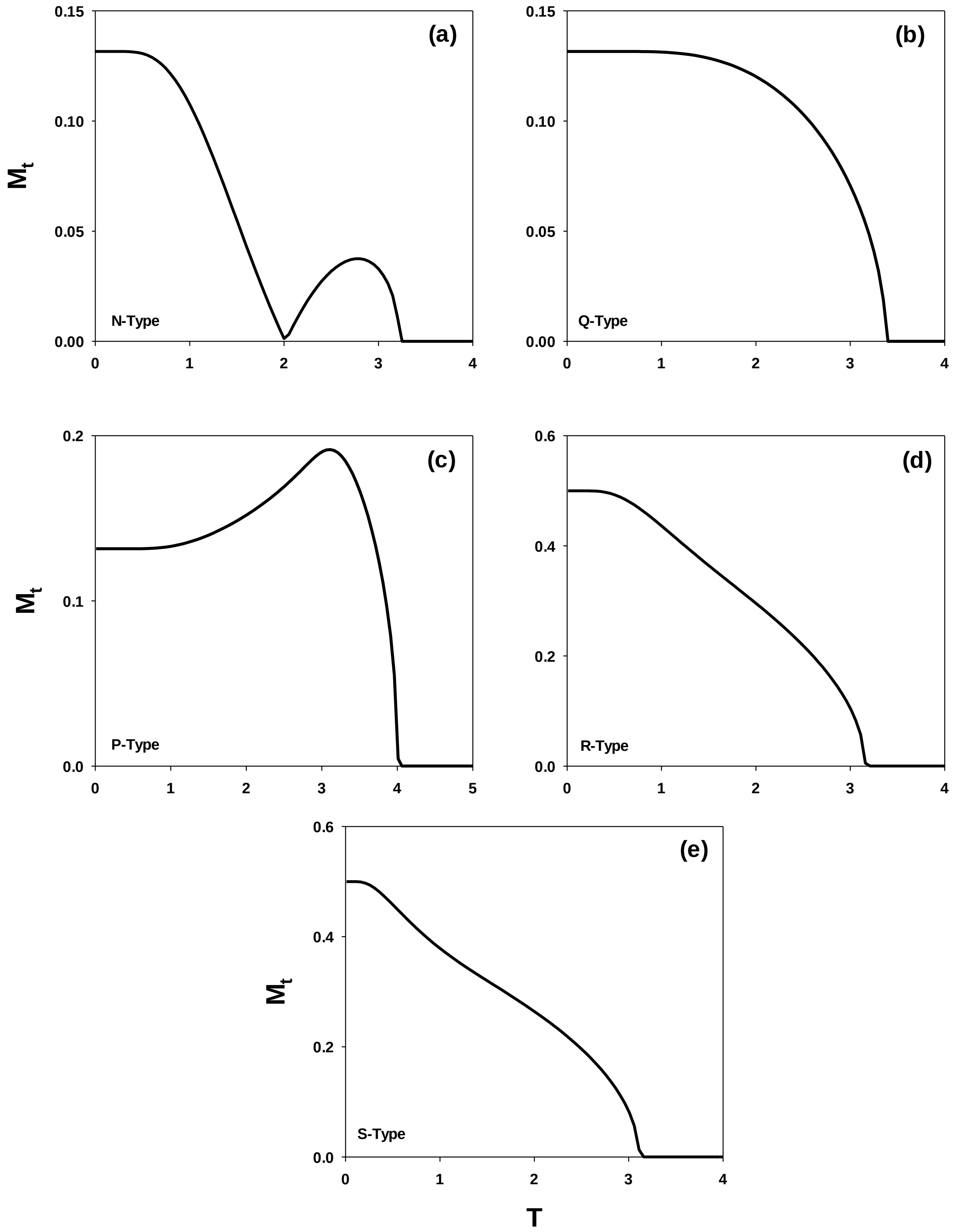

(a)
N-Type
(b)
Q-Type
(c)
P-Type
(d)
R-Type
(e)
S-Type
M_t
T
0.15
0.10
0.05
0.00
0.2
0.1
0.0
0.6
0.4
0.2
0
1
2
3
4
5

Fig. 3

Fig. 4

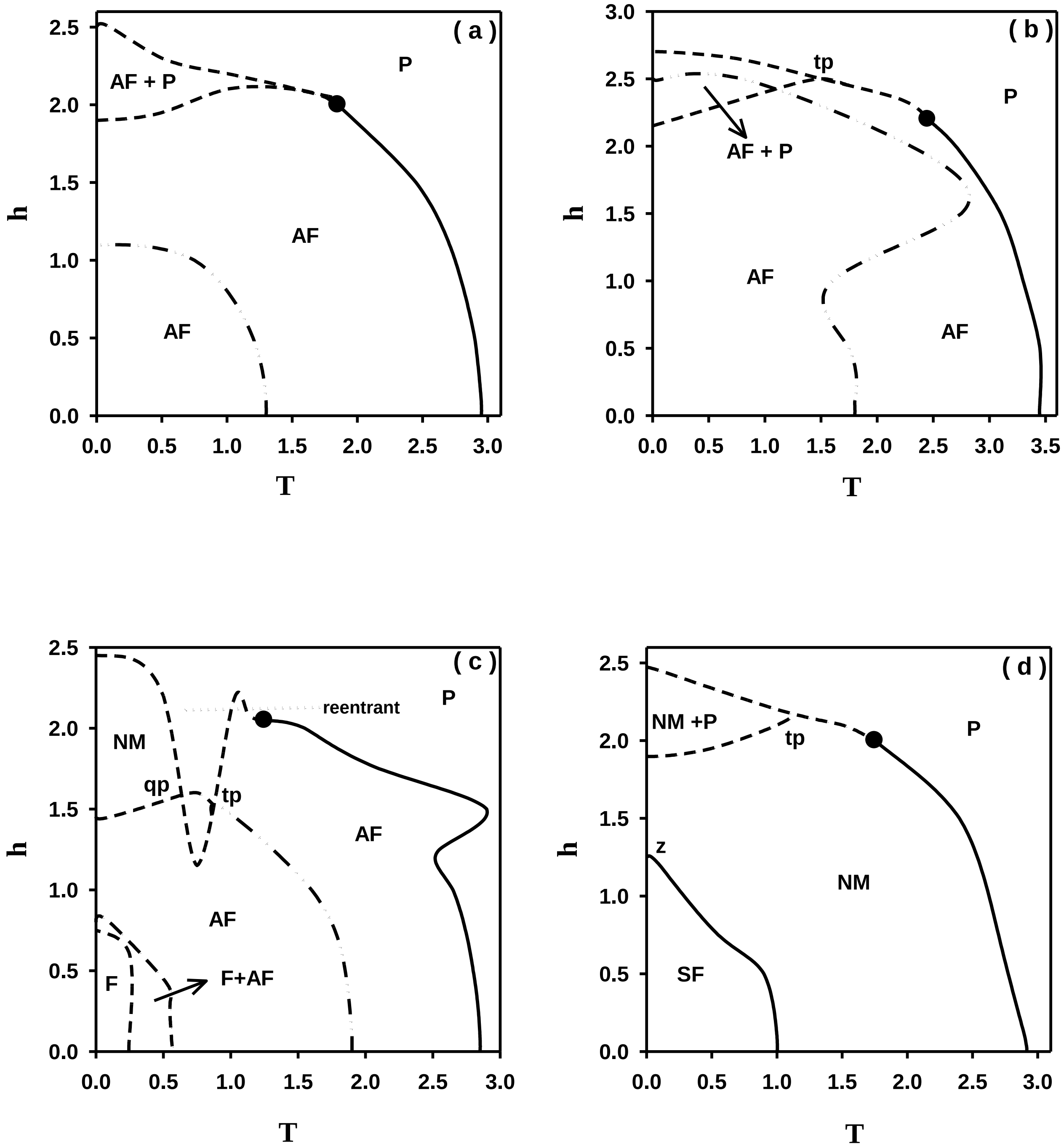
( a )
AF + P
P
AF
AF
h
T
( b )
tp
P
AF + P
AF
AF
h
T
( c )
reentrant
P
NM
qp
tp
AF
AF
F
F+AF
h
T
( d )
NM +P
tp
P
z
NM
SF
h
T

**Fig. 4**

Fig. 5

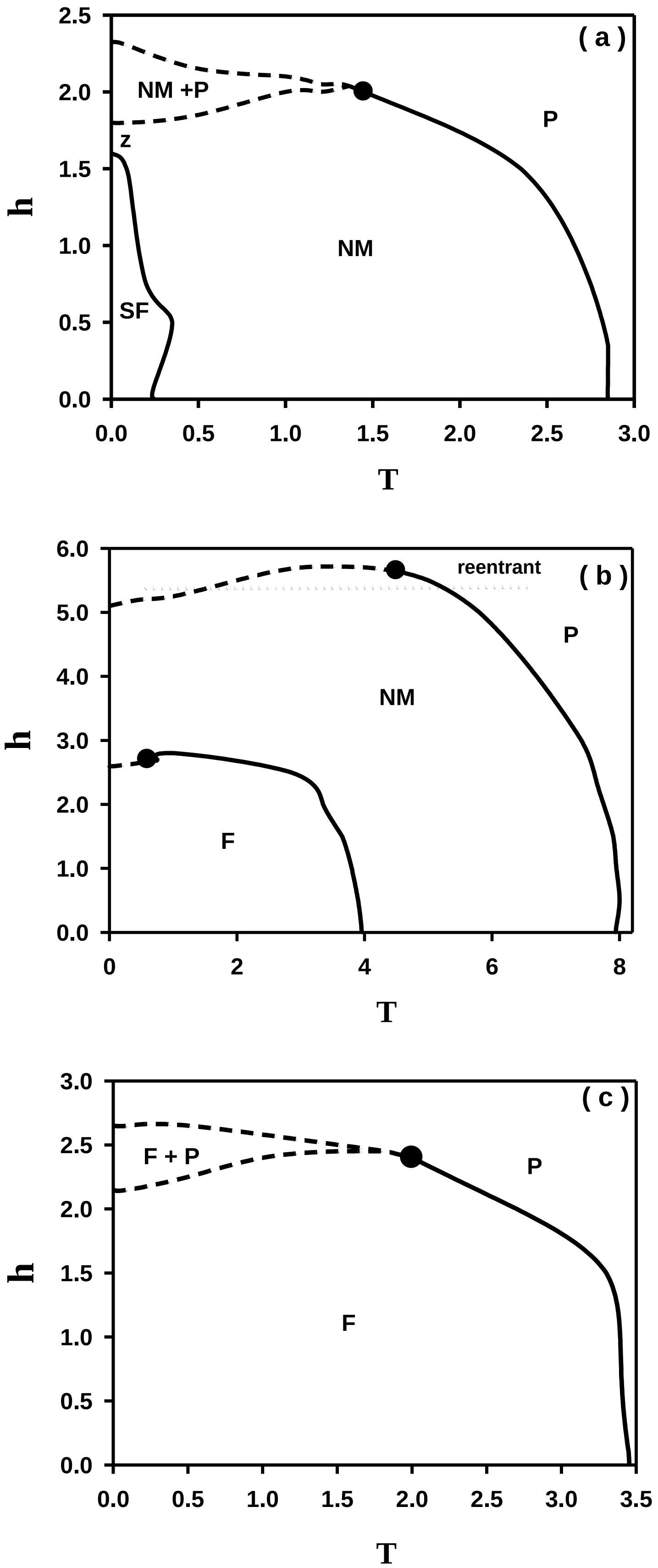
( a )
NM +P
P
z
NM
SF
h
T
2.5
2.0
1.5
1.0
0.5
0.0
0.0
0.5
1.0
1.5
2.0
2.5
3.0
( b )
reentrant
P
NM
F
6.0
5.0
4.0
3.0
2.0
1.0
0.0
0
2
4
6
8
( c )
F + P
P
F
3.0
2.5
2.0
1.5
1.0
0.5
0.0
0.0
0.5
1.0
1.5
2.0
2.5
3.0
3.5

**Fig. 5**